\documentclass[12pt]{iopart}
\usepackage{amsmath}
\usepackage{iopams}
\usepackage{cite}
\usepackage{graphicx}
\begin{document}

\title[Adversarial attacks on prisoner's dilemma in complex networks]{Steering cooperation: Adversarial attacks on prisoner's dilemma in complex networks}

\author{Kazuhiro Takemoto}

\address{Department of Bioscience and Bioinformatics, Kyushu Institute of Technology, Iizuka, Fukuoka 820-8502, Japan}
\ead{takemoto@bio.kyutech.ac.jp}
\vspace{10pt}
\begin{indented}
\item[]\today
\end{indented}

\begin{abstract}
This study examines the application of adversarial attack concepts to control the evolution of cooperation in the prisoner's dilemma game in complex networks. Specifically, it proposes a simple adversarial attack method that drives players' strategies towards a target state by adding small perturbations to social networks. The proposed method is evaluated on both model and real-world networks. Numerical simulations demonstrate that the proposed method can effectively promote cooperation with significantly smaller perturbations compared to other techniques. Additionally, this study shows that adversarial attacks can also be useful in inhibiting cooperation (promoting defection). The findings reveal that adversarial attacks on social networks can be potent tools for both promoting and inhibiting cooperation, opening new possibilities for controlling cooperative behavior in social systems while also highlighting potential risks.
\end{abstract}

%
\vspace{2pc}
\noindent{\it Keywords}: prisoner's dilemma, evolutionary game, complex networks, adversarial attacks
%
%
%
%

\section{Introduction}
The prisoner's dilemma game is a fundamental model for investigating the evolution of cooperation among self-interested individuals \cite{SZABO200797,PERC20171,ohtsuki2006replicator}.
This game presents a social dilemma where mutual cooperation maximizes collective benefit, yet defection is each player's rational choice.
This paradigm reflects universal challenges in various real-world scenarios, such as resource management, environmental conservation, and collective action, making the search for solutions highly significant.

Research on the prisoner's dilemma game in complex networks initially focused on lattice networks with uniform connectivity, revealing that spatial structure promotes cooperation evolution \cite{nowak1992evolutionary,szabo1998evolutionary,nowak1993spatial}.
Advancements in network science \cite{albert2002statistical,Takemoto2012_book} have uncovered non-trivial connectivity patterns in real-world social networks, such as small-world topology \cite{watts1998collective} and heterogeneous or scale-free connectivity \cite{barabasi1999emergence,barabasi2003scale}.
Studies have explored the impact of these realistic network structures on cooperation evolution.
Heterogeneous connectivity often promotes cooperation due to hubs \cite{santos2005scale,santos2006graph,wu2007evolutionary}, although some studies challenge this view \cite{szolnoki2008towards,gracia2012heterogeneous}.
Conversely, small-world networks tend to inhibit emergence of cooperation due to shortcuts or long-range interactions \cite{watts1998collective,abramson2001social,kim2002dynamic}.

The critical role of influential nodes in maintaining cooperation within networks has been highlighted.
Studies have shown that while cooperation is resilient to random node removal, it is vulnerable to targeted removal of high-degree nodes \cite{perc2009evolution}.
Further investigations have demonstrated that considering a node's extended neighborhood (collective influence) can be more effective in determining its impact on cooperation than simply its degree \cite{szolnoki2016collective}.
These findings underscore the importance of key nodes in fostering cooperative behavior.

Coevolutionary approaches, allowing for the simultaneous evolution of strategies and network properties \cite{perc2010coevolutionary}, have expanded our understanding of cooperation dynamics. In this context, studies have examined the impact of link weights on cooperation evolution. Both heterogeneity \cite{du2008evolutionary,meng2016interdependency,iwata2016heterogeneity} and adaptive changes in link weights \cite{cao2011evolution,li2019reputation,liu2020link} have been shown to promote cooperation.
These approaches, particularly promising in online social networks, suggest that cooperation can be enhanced by strategically adjusting interaction strengths between individuals without direct user intervention.

However, while providing valuable insights, these methods often require substantial network alterations, making practical implementation costly and challenging. To address this limitation, this study proposes leveraging the concept of adversarial attacks, originally developed for neural networks \cite{DBLP:journals/corr/SzegedyZSBEGF13,DBLP:journals/corr/GoodfellowSS14}.
The fundamental principle of adversarial attacks is that carefully designed small perturbations can dramatically influence system behavior when strategically applied to system parameters.
The application of this concept to network dynamics has been explored in voter models \cite{chiyomaru2022adversarial,chiyomaru2023mitigation}, where minimal adjustments to network parameters (particularly to link weights) effectively guide collective behavior toward desired states.

In the context of the prisoner's dilemma game, such an approach offers a promising direction for controlling cooperation.
While players make decisions based on local payoffs and neighbors' strategies, strategic perturbations to their interaction strengths through link weights could guide the overall system toward cooperative states.
This framework enables the promotion of cooperation through minimal network modifications, potentially offering a more practical approach than traditional methods requiring large-scale structural changes.

Based on this idea, this study proposes a method to promote cooperation evolution in the prisoner's dilemma game in complex networks using adversarial attacks.
The method adds small, strategically generated perturbations to link weights to foster cooperation, while also demonstrating the approach's ability to inhibit it.
The proposed method is numerically evaluated across various networks, including representative models and real-world social networks.
Additionally, the performance of this adversarial attack approach is compared with existing methods for link weight adjustment, demonstrating its potential advantages in promoting cooperation.
Furthermore, the potential negative aspects and ethical implications of this approach are also examined.

\section{Methods}
\subsection{Prisoner's dilemma game in complex networks}
This study employs a prisoner's dilemma game in a network of $N$ nodes (players), based on settings used in previous research \cite{SZABO200797,santos2005scale,santos2006graph}.
At time $t$, each player adopts either a cooperative strategy (C; $\boldsymbol{\sigma}_i(t)=(1, 0)$) or a defective strategy (D; $\boldsymbol{\sigma}_i(t)=(0, 1)$), and gains payoffs by playing the game with neighboring players.
Following previous studies \cite{du2008evolutionary,iwata2016heterogeneity,buesser2012supercooperation,li2019reputation,liu2020link,wen2009evolutionary,zhai2010effective,li2017co} on the prisoner's dilemma game in weighted networks, which focus on adjusting link weights, the total payoff for player $i$ at time $t$ is calculated as follows:

\begin{equation}
    P_i(t) = \sum_{j=1}^NA_{ij}\left[\boldsymbol{\sigma}_i(t) \boldsymbol{M} \boldsymbol{\sigma}_j^{\top}(t)\right]
    \label{eq:compute_total_payoff}
\end{equation}

Here, $A_{ij}$ represents an element of the weighted adjacency matrix $\boldsymbol{A}$ of the network, indicating player $j$'s influence on player $i$.
For simplicity, this study considers complex networks where the existence of a connection between players is bidirectional, but the strength of influence can differ in each direction.
Initially, uniform link weights of 1 are set (i.e., each player exerts equal influence on connected neighbors).
Specifically, if players $i$ and $j$ can influence each other, the initial weights are set as $A_{ij} = A_{ji} = 1$, while if no relationship exists between players, the corresponding matrix elements are set to zero ($A_{ij} = A_{ji} = 0$).
Self-loops are excluded (i.e., $A_{ii}=0$ for $i=1,\dots,N$).
After applying adversarial attacks, while the existence of connections remains bidirectional, the influence strengths can become asymmetric ($A_{ij} \neq A_{ji}$), reflecting different strengths of influence in each direction.
These connections with initially uniform weights form the baseline network structure for the analysis of adversarial attacks.

$\boldsymbol{M}$ is the game's payoff matrix.
This study employs a specific variant of the prisoner's dilemma game known as the weak or boundary game, where the payoff matrix is simplified as follows:
\begin{equation}
    \boldsymbol{M}= \left( {\begin{array}{cc} 1 & 0 \\b & 0 \end{array} } \right). 
    \nonumber
\end{equation}
Here $b$ represents the advantage of defectors over cooperators, constrained by $1 < b < 2$ in  this version of the game \cite{nowak1992evolutionary,nowak1993spatial,santos2005scale,santos2006graph}.
This simplified version differs from the general (strong) prisoner's dilemma game, where payoffs satisfy $T > R > P > S$ and $2R > T + S$ ($T$ is the temptation to defect, $R$ is the reward for mutual cooperation, $P$ is the punishment for mutual defection, and $S$ is the sucker's payoff).
In our simplified version, the payoffs are set as $R = 1$, $P = S = 0$, and $T = b$.

This boundary game represents a specific case where there is no stag hunt-type dilemma (as $P=S$) but maintains a chicken-type dilemma with strength $(T-R)/(R-P)=b-1$ \cite{wang2015universal}.
While this formulation has been widely used in previous studies \cite{nowak1992evolutionary,nowak1993spatial,santos2005scale,santos2006graph,li2017co,abramson2001social,wang2012effects,devlin2009cooperation,szabo2009cooperation,szabo1998evolutionary,liu2020evolution,wen2009evolutionary,meng2016interdependency,du2008evolutionary,szolnoki2008towards,wu2007evolutionary,cao2011evolution,li2020evolution}, it is important to acknowledge that a more general form, known as the donor-recipient game, is often considered the standard template in theoretical biology \cite{wang2015universal}.
In this game, cooperation involves a cost $c$ incurred by the donor to provide a benefit $b$ to the recipient ($T=b$, $R=b-c$, $P=0$, $S=-c$), resulting in equal dilemma strengths $c/(b-c)$.
Our choice of the boundary game allows us to focus on a single parameter $b$ while preserving the fundamental social dilemma where rational individual choices lead to defection despite mutual cooperation yielding collectively better outcomes \cite{nowak1992evolutionary,cao2011evolution,roy2023time}.

Player $i$ determines their strategy for the next time step ($t+1$) based on their total payoff and those of their neighbors (strategy update).
The basic idea is that players tend to imitate the strategies of more successful neighbors (i.e., those with higher payoffs).
Specifically, player $i$ imitates the strategy of player $j$, selected with a probability proportional to the link weight ($A_{ij}/\sum_h A_{ih}$), according to the following Fermi rule \cite{zhi2006prisoner,buesser2012supercooperation,liu2020link,liu2020evolution}:
\begin{equation}
    \phi_{ij}=\frac{1}{1+\exp\left[\left\{P_i(t)-P_j(t)\right\}/K\right]}.
    \label{eq:transition_probability}
\end{equation}
Here, $K$ represents environmental noise in the strategy adoption process, including factors such as irrationality and errors in decision making.
A higher value of $K$ indicates more randomness in strategy adoption.
The effect of noise $K$ has been studied in detail in previous research \cite{szabo2005phase,vukov2006cooperation,szolnoki2009topology}, showing that moderate levels of noise can facilitate the emergence and maintenance of cooperation.
Following previous studies, $K$ is set to 0.1 \cite{szabo1998evolutionary,du2008evolutionary,liu2020link,liu2020evolution,meng2016interdependency,wu2007evolutionary}.
This value represents a moderate level of noise that allows for some randomness in strategy adoption while still maintaining the influence of payoff differences.
It has been confirmed that using an alternative update rule, such as the proportional rule employed in other studies \cite{santos2005scale,santos2006graph,wu2007evolutionary,li2019reputation}, does not alter the conclusions of this study.

Players' strategies are updated synchronously, and this update is repeated $t_{\max}$ times.
It has been verified that asynchronous strategy updates do not change the study's conclusions.
Previous research has shown that whether strategy updates are synchronous or asynchronous does not qualitatively affect the results \cite{santos2006graph,hauert2004spatial}.

\subsection{Adversarial attacks}
Following the approach of adversarial attacks on voter model dynamics \cite{chiyomaru2022adversarial,chiyomaru2023mitigation}, this study considers adversarial attacks that drive players' strategies towards a target state in the next time step (i.e., $t+1$). While the network initially starts with uniform link weights, these weights are modified over time as part of the attack strategy.

Based on these previous studies \cite{chiyomaru2022adversarial,chiyomaru2023mitigation}, this study employs an energy function $E$ that quantifies the distance between the current system state and the desired target state.
For computational convenience, player $i$'s strategy at time $t$ is defined as $x_i(t)=(+1, -1)\boldsymbol\sigma^{\top}_i(t)$.
Thus, $x_i(t) = +1~(-1)$ if player $i$ chooses strategy C (D) at time $t$.
Let $x_i^* = \{+1, -1\}$ be the target state for players, and the attack aims to achieve $x_i(t+1)=x_i^*$ for $i=1,\dots, N$.
The attacks are applied by minimizing the energy $E=-N^{-1}\sum_{i=1}^N x^*_i x_i(t+1)$, which represents the negative correlation coefficient between the observed state and target state, effectively capturing the system's deviation from the desired state.

To promote cooperation, $x^*_i=+1$ is set for $i=1,\dots,N$.
Consequently, the energy $E$ to be minimized becomes:
\begin{equation}
    E=-\frac{1}{N}\sum_{i=1}^N x_i(t+1).
    \label{eq:energy}
\end{equation}

To minimize $E$, the study considers temporally varying link weights.
Specifically, the gradient descent method is used to add perturbations to $\boldsymbol{A}$ at each time step.
Assuming that link weights between unconnected node pairs in the original network cannot be changed, perturbations are added to the link weights of node pairs $i$ and $j$ where $A_{ij}\neq 0$ at time $t$ as follows:
\begin{equation}
    A_{ij}^{*}(t) =  A_{ij} - \epsilon \frac{\partial E}{\partial A_{ij}}.
    \nonumber
\end{equation}
Here, $\epsilon$ is a small value.
Note that setting $\epsilon>0$ results in adversarial attacks to promote cooperation (targeting C), while $\epsilon<0$ leads to attacks inhibiting cooperation or, equivalently, promoting defection (targeting D).

However, the gradient $\partial E/\partial A_{ij}$ cannot be directly (analytically) obtained.
Drawing from previous studies \cite{chiyomaru2022adversarial,mizutaka2023crossover}, the following mean-field time evolution of player $i$'s strategy in the prisoner's dilemma game is considered:
\begin{equation}
    x_i(t+1)=\sum_{j=1}^N\frac{A_{ij}}{\sum_{h=1}^N A_{ih}}\left[\phi_{ij}x_j(t)+(1-\phi_{ij})x_i(t)\right]
    \label{eq:strategy_evolution}
\end{equation}

Substituting Equation (\ref{eq:strategy_evolution}) into Equation (\ref{eq:energy}) yields $E$ s a function of $A_{ij}$.
The gradient can then be derived through straightforward differentiation with respect to $A_{ij}$, which results in:
\begin{equation}
    \frac{\partial E}{\partial A_{ij}}=\frac{1}{N\left(\sum_{h=1}^N A_{ih}\right)^2}\left[A_{ij}\phi_{ij}\{x_i(t)-x_j(t)\}\sum_{\substack{h=1 \\ h\neq j}}^N A_{ih} - \sum_{\substack{h=1 \\ h\neq j}}^N A_{ih}\phi_{ih}\{x_i(t) - x_h(t)\}\right].
    \label{eq:gradient}
\end{equation}

The effect of $\epsilon$ can be intuitively understood through this gradient: the gradient descent method modifies link weights to minimize the energy $E$.
This modification tends to strengthen the influence of successful players adopting the target strategy on their neighbors while weakening the influence of players with the opposite strategy.
When $\epsilon>0$, this mechanism promotes the spread of cooperative strategies by enhancing cooperators' influence, while $\epsilon<0$ leads to the opposite effect, strengthening defectors' influence instead.
However, as shown in Equation (\ref{eq:gradient}), the actual mechanism is more complex, involving intricate interactions between network structure, payoffs, and strategy adoption probabilities.
These effects are achieved through small perturbations that do not significantly alter the underlying network structure.

However, directly using this gradient for adversarial attacks is not practical as the exact values of $\phi_{ij}$ are unknown in real scenarios.
Therefore, following previous research \cite{chiyomaru2022adversarial,chiyomaru2023mitigation}, this study takes a more practical approach by utilizing the optimal maximum norm constrained perturbation (perturbation based on the sign of the gradient) \cite{DBLP:journals/corr/GoodfellowSS14}.
While this approach does not capture the exact magnitude of the gradient, it can still estimate its direction.
Since $A_{ij} \geq 0$, $x_i(t)\in \{+1,-1\}$, and $N>0$, the sign of the gradient remains consistent even if $\phi_{ij}$ is replaced with another $\phi'_{ij}$ that monotonically increases with respect to $P_i(t) - P_j(t)$.
This property allows the determination of whether each link weight should be increased or decreased, even without knowing the exact values of $\phi_{ij}$.
Thus, the adversarial attacks are implemented by modifying each element of $\boldsymbol{A}$ based on the estimated sign of the gradient:
\begin{equation}
    A_{ij}^{\mathrm{adv}}(t) = A_{ij} - \epsilon \times \mathrm{sign}\left(\frac{\partial E}{\partial A_{ij}}\right).
    \nonumber
\end{equation}
Using the sign of the gradient in this way also has the advantage of controlling the strength of the perturbations with $\epsilon$.

The adversarial attacks on the prisoner's dilemma game are primarily executed by replacing $\boldsymbol{A}$ with $\boldsymbol{A}^{\mathrm{adv}}$ in Equation (\ref{eq:compute_total_payoff}).
This change indirectly affects Equation (\ref{eq:transition_probability}) through the modified payoffs.

\subsection{Comparison with other methods}
To assess the efficacy of the proposed adversarial attacks, comparisons are made with random attacks and Li et al.'s link weight adjustment method \cite{li2019reputation}.

Random attacks are introduced as a critical control condition, not as a competitive method.
Their primary purpose is to provide a baseline for evaluating the significance of gradient consideration in the adversarial attacks.
By comparing the effects of adversarial attacks to those of random perturbations, the impact of the gradient-based strategy can be isolated and quantified.
These attacks involve adding perturbations to $\boldsymbol{A}$ according to the following formula: $A_{ij}^{\mathrm{rnd}}(t) = A_{ij} + \epsilon \times s$ for all pairs $(i,j)$ where $A_{ij} \neq 0$.
In this context, $s$ represents a random variable sampled from the set $\{-1,+1\}$.
The implementation of the random attacks primarily involves substituting $\boldsymbol{A}$ with $\boldsymbol{A}^{\mathrm{rnd}}$ in Equation (\ref{eq:compute_total_payoff}).

While random attacks are not expected to have a consistent directional effect, they may show some impact due to the stochastic nature of the process and the complex dynamics of the network. Any observed effect in random attacks underscores the network's sensitivity to perturbations and highlights the potential of targeted strategies.

Li et al.'s method is employed as a representative link weight adjustment technique for promoting cooperation.
This selection is justified by its status as a state-of-the-art approach and, analogous to the proposed method, its capacity to regulate the magnitude of link weight modifications (perturbation strength), thus facilitating a fair comparison.
In Li et al.'s method, an adaptive adjustment is implemented by comparing a player's total payoff with the mean total payoff of their neighboring players. The link weight is increased (decreased) by a unit amount $d$ if the player's total payoff exceeds (falls below) the average.
Specifically, this method initializes $A_{ij}^{\mathrm{Li}}(0) \leftarrow A_{ij}$ and, assuming $|V(i)|>0$ for $i = 1,\dots,N$, where $V(i)$ denotes the set of the neighbors of node $i$, iteratively updates link weights of node pairs $i$ and $j$ where $A_{ij}\neq 0$ according to the following equation:
\begin{equation}
    A_{ij}^{\mathrm{Li}}(t+1)\leftarrow \mathrm{clip}_{A{ij}\pm\epsilon}\left[A_{ij}^{\mathrm{Li}}(t)+d\times \mathrm{sign}\left(P_i(t)-\frac{\sum_{j\in V(i)}P_j(t)}{|V(i)|}\right)\right].
    \nonumber
\end{equation}
Here, $\mathrm{clip}_{A_{ij}\pm\epsilon}(x)$ is a function that constrains $x$ to the interval $[A_{ij} - \epsilon, A_{ij} + \epsilon]$.
Note that $0\leq \epsilon< A_{ij}$ (given the configuration of $\boldsymbol{A}$ in this study, $0\leq \epsilon < 1$).
In accordance with previous research, $d$ is set to 0.01.
It is important to note that due to the clipping function and the condition $0\leq \epsilon< A_{ij}$, non-zero weights will never become negative or zero after applying perturbations.
The prisoner's dilemma game based on this method is primarily executed by substituting $\boldsymbol{A}$ with $\boldsymbol{A}^{\mathrm{Li}}$ in Equation (\ref{eq:compute_total_payoff}).

These comparisons allow us to demonstrate both the necessity of a targeted approach (versus random perturbations) and the effectiveness of our gradient-based method against existing strategies.

\section{Results}
To evaluate the effect of adversarial attacks on the prisoner's dilemma game, simulations were first conducted on complex networks generated from three representative network models: the Erd\H{o}s--R\'enyi (ER) model \cite{albert2002statistical,Takemoto2012_book}, Barab\'asi--Albert (BA) \cite{albert2002statistical,barabasi1999emergence}, and Watts--Strogatz (WS) models \cite{watts1998collective}.

The ER model, widely used in network science, generates random networks by establishing links between $L$ randomly selected node pairs from all possible pairs.
In this model, the node degree $k$ follows a Poisson distribution with $\langle k \rangle = 2L/N$, where $\langle k \rangle$ represents the average degree of the network.
However, real-world social networks exhibit nonrandom structures, characterized by highly clustered subnetworks and heterogeneous (power-law-like) degree distributions \cite{albert2002statistical,barabasi1999emergence,barabasi2003scale}.
Thus, BA and WS models were considered.
The BA model generates scale-free random networks by iteratively adding new nodes and connecting them to $m$ existing nodes using a preferential attachment mechanism.
This process results in a degree distribution $P(k)$ that follows a power law [$P(k)\propto k^{-3}$]. It should be noted that $\langle k \rangle = 2m$ for $N \gg 0$.
The WS model generates small-world networks by randomly rewiring links in a one-dimensional lattice, where each node initially has $k$ ($=\langle k \rangle$) neighbors, with probability $p_{\mathrm{WS}}$. The resulting networks exhibit higher clustering coefficients than those expected from ER networks. In this study, $p_{\mathrm{WS}} = 0.05$, following \cite{chiyomaru2022adversarial}.

In the simulations, unless otherwise specified, networks with $N = 1000$ nodes and an average degree of $\langle k \rangle = 8$ were used.
It has been confirmed that using different values of $N$ (ranging from 200 to 4000) or $\langle k \rangle$ (ranging from 4 to 32) does not qualitatively affect the results.
The initial state ($t = 0$) consisted of an equal number of cooperators and defectors, and the simulation time $t_{\max}$ was set to 1000 steps.
The proportion of cooperators $\rho$ in the equilibrium state was calculated by averaging over the last 100 steps.
It has been verified that this $t_{\max}$ setting is sufficient for $\rho$ to reach equilibrium, and that moderate changes in the number of steps used for averaging do not affect the results.
The structure of model networks and the initial distribution of cooperators and defectors within the networks depend on random number generation.
To minimize the impact of individual network structures and initial states' randomness, and to more accurately evaluate the effect of adversarial attacks, the value of $\rho$ is reported as the average of results obtained from 100 independent trials.

\begin{figure}[htbp]
\begin{center}
\includegraphics[width=78mm]{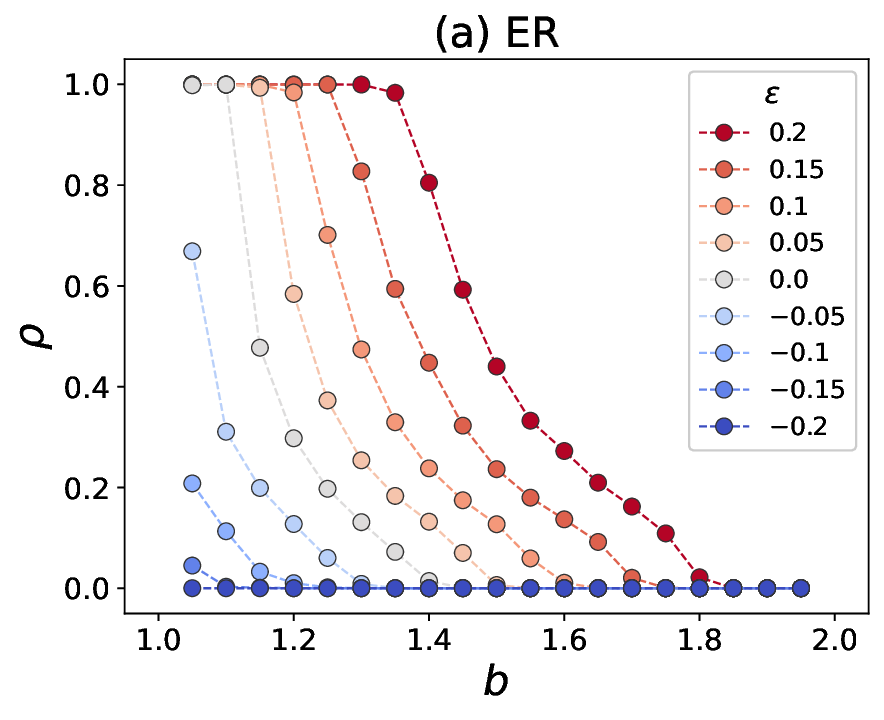}
\includegraphics[width=78mm]{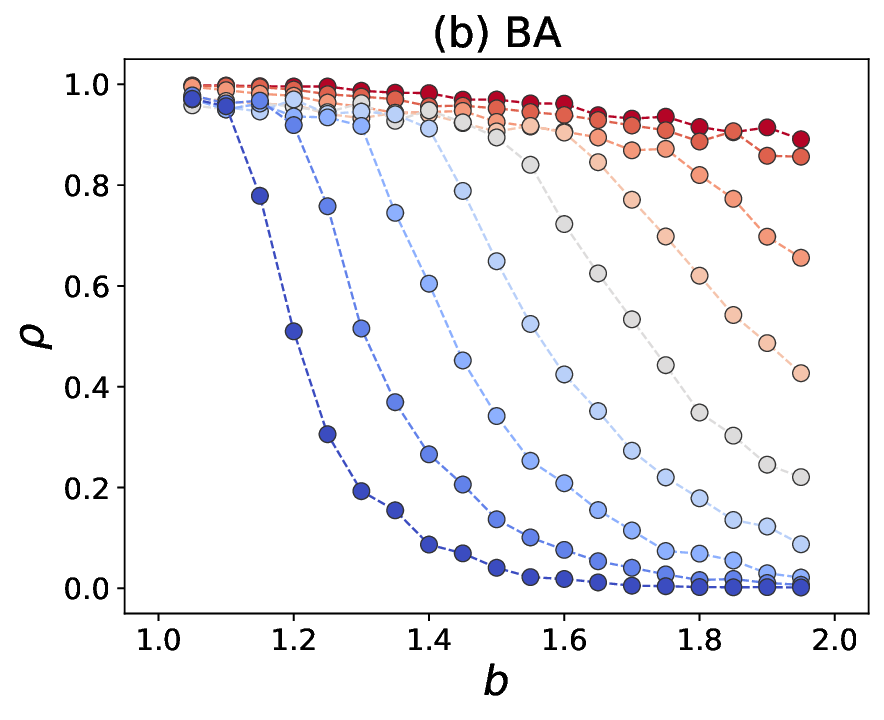} \\ 
\includegraphics[width=78mm]{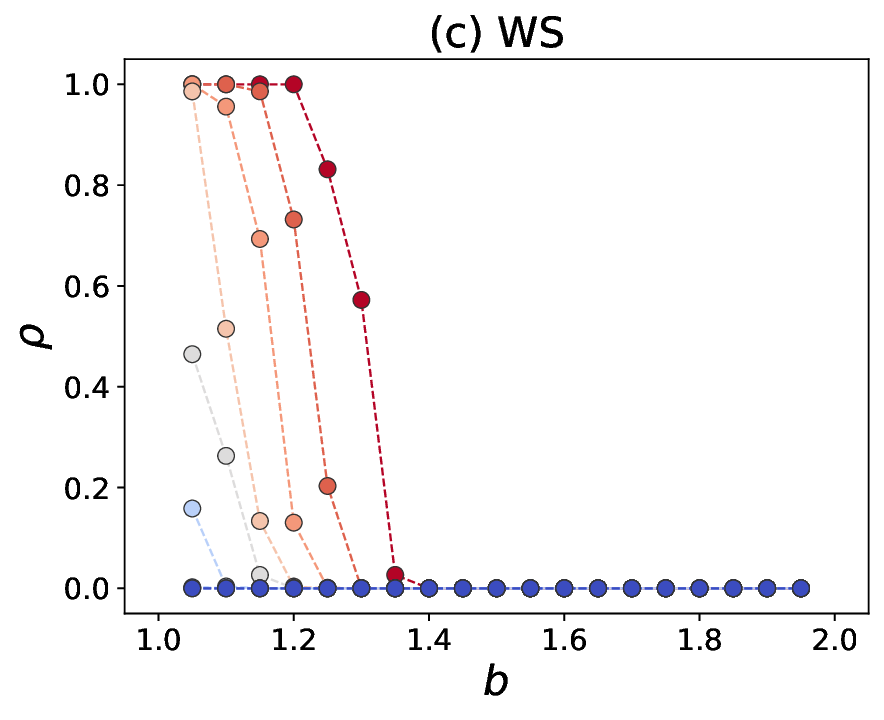}
\end{center}
\caption{\label{fig:rho_vs_b_wrt_eps_model} Line plots of the proportion of cooperators $\rho$ versus the advantage of defectors $b$ in (a) Erd\H{o}s--R\'enyi (ER) random networks, (b) Barab\'asi--Albert (BA) scale-free networks, and (c) Watts--Strogatz (WS) small-world networks, showing that adversarial attacks ($\epsilon$) can both promote ($\epsilon > 0$) and inhibit ($\epsilon < 0$) cooperation.}
\end{figure}

Figure \ref{fig:rho_vs_b_wrt_eps_model} demonstrates that adversarial attacks can both promote and inhibit cooperation across different network models (ER, BA, and WS).
For promoting cooperation ($\epsilon > 0$), $\rho$ maintains higher values even for large $b$ compared to the case without attacks ($\epsilon = 0$).
This effect is observed in all network types, with ER networks (Figure \ref{fig:rho_vs_b_wrt_eps_model}a) showing a substantial increase in $\rho$ from about $0.2$ ($\epsilon = 0$) to nearly $1.0$ ($\epsilon = 0.15$) at $b = 1.25$.
The range of b where cooperation dominates ($\rho > 0.5$) is significantly broadened, as seen in the extension of the upper limit of $b$ in ER networks from approximately $1.15$ ($\epsilon = 0$) to around $1.5$ ($\epsilon = 0.2$).

Similar promotion of cooperation is observed in BA and WS networks (Figures \ref{fig:rho_vs_b_wrt_eps_model}b and \ref{fig:rho_vs_b_wrt_eps_model}c), albeit with some differences due to their unique structural properties.
In BA networks, which inherently promote cooperation, the effect is particularly pronounced.
For example, at $b = 1.95$, $\rho$ increases from about $0.2$ ($\epsilon = 0$) to nearly $0.9$ ($\epsilon = 0.2$), with the upper limit of $b$ for dominant cooperation extending beyond $1.95$ when $\epsilon \geq 0.1$.
WS networks, despite their tendency to inhibit cooperation, also show significant improvements.
At $b = 1.15$, $\rho$ increases from roughly $0.05$ ($\epsilon = 0$) to approximately $1.0$ ($\epsilon = 0.15$), with the upper limit of $b$ for dominant cooperation increasing from about $1.05$ ($\epsilon = 0.0$) to around $1.3$ ($\epsilon = 0.2$).

The effect of adversarial attacks is also remarkable for the inhibition of cooperation ($\epsilon < 0$) (Figure~\ref{fig:rho_vs_b_wrt_eps_model}).
While $\rho$ decreases with increasing perturbation strength $|\epsilon|$ in adversarial attacks, the decrease due to random attacks is limited across all network types.
In ER networks, at $b = 1.1$, $\rho$ decreases from nearly $1.0$ ($\epsilon = 0$) to approximately $0.1$ ($\epsilon = -0.1$), with the upper limit of $b$ for dominant cooperation falling below $1.05$ when $\epsilon = -0.1$. 
BA networks, despite their inherent cooperation-promoting nature, show a significant reduction in cooperation under adversarial attacks.
For instance, at $b = 1.5$, $\rho$ decreases from around $0.9$ ($\epsilon = 0$) to roughly $0.05$ ($\epsilon = -0.2$), with the upper limit of $b$ for dominant cooperation decreasing to about $1.2$.
In WS networks, the effect is even more pronounced, with $\rho$ dropping to approximately $0.0$ at $b = 1.05$ when $\epsilon \leq -0.1$.

\begin{figure}[htbp]
\begin{center}
\includegraphics[width=77mm]{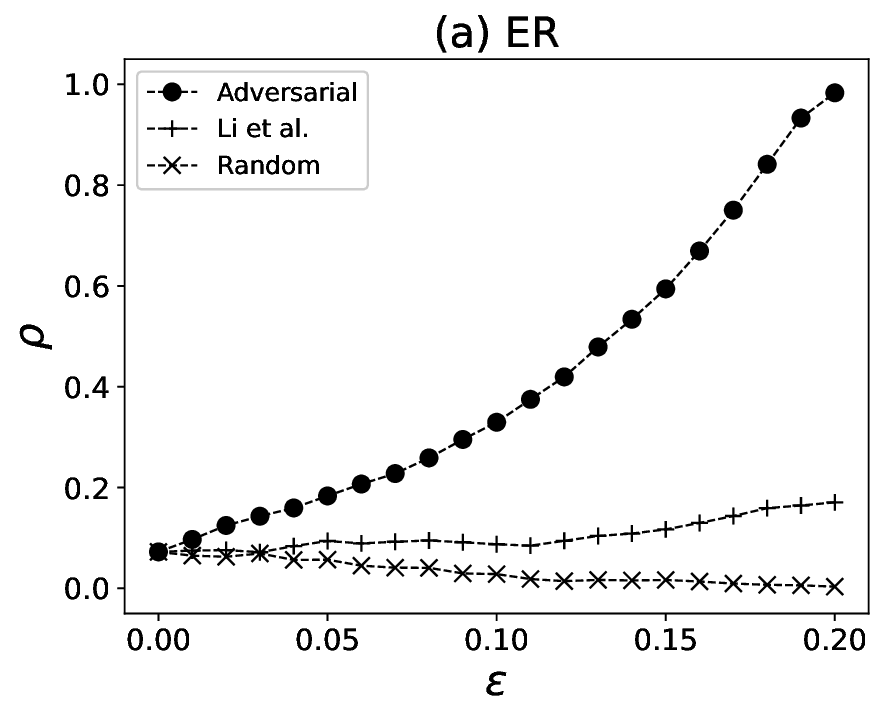} 
\includegraphics[width=77mm]{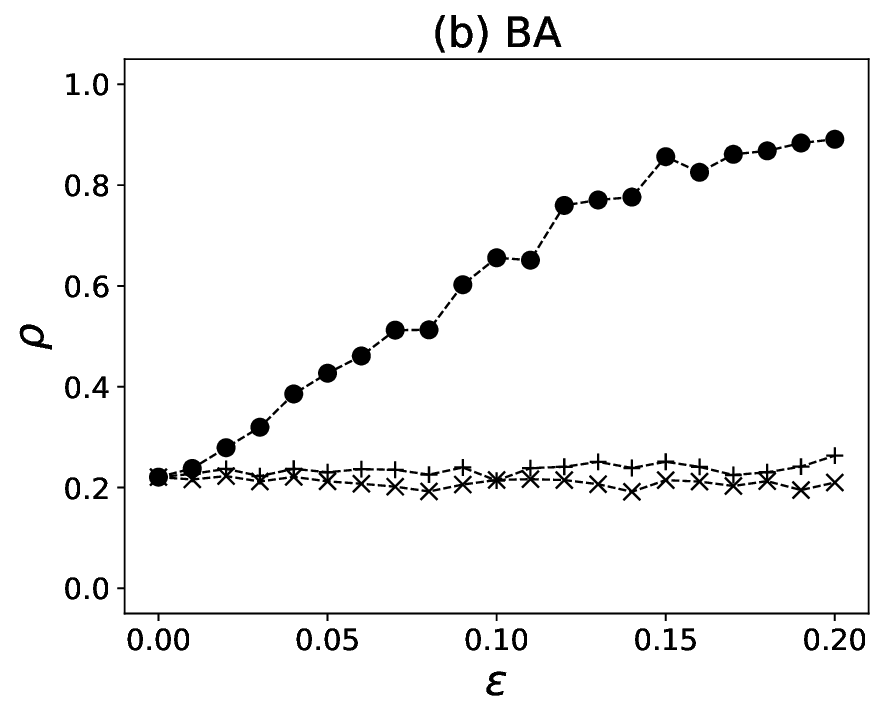} \\ 
\includegraphics[width=77mm]{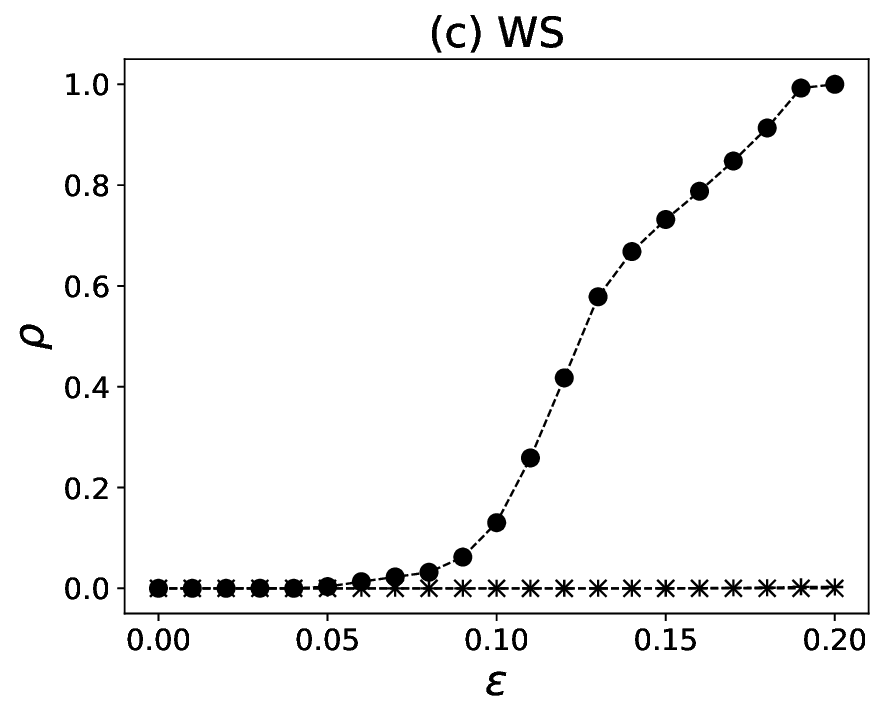} 
\end{center}
\caption{\label{fig:rho_vs_eps_coop_model} Line plots of the proportion of cooperators $\rho$ versus perturbation strength $\epsilon$ for adversarial attacks promoting cooperation ($\epsilon > 0$) in (a) Erd\H{o}s--R\'enyi (ER) random networks, (b) Barab\'asi--Albert (BA) scale-free networks, and (c) Watts--Strogatz (WS) small-world networks, showing that adversarial attacks promote cooperation substantially more effectively than random attacks and Li et al.'s method The advantage of defectors $b$ is 1.35, 1.95, and 1.2 for (a), (b), and (c), respectively. These $b$ values are chosen as the maximum values for which $\rho \leq 0.1$ when $\epsilon = 0$ in Fig. \ref{fig:rho_vs_b_wrt_eps_model}, or set to 1.95 if no such value exists within $1 < b < 2$ (as in Fig. \ref{fig:rho_vs_b_wrt_eps_model}b).}
\end{figure}

The superiority of adversarial attacks becomes evident when compared to random attacks and Li et al.'s method (Figure~\ref{fig:rho_vs_eps_coop_model}).
For promoting cooperation ($\epsilon > 0$), adversarial attacks show a rapid increase in $\rho$ with increasing $\epsilon$, while random attacks fail to increase $\rho$ in any of the model networks examined. The ineffectiveness of random attacks, serving as a control condition, highlights the importance of the gradient-based approach in our adversarial attacks. This comparison demonstrates that the targeted nature of the gradient consideration, rather than mere network perturbation, is crucial for effectively promoting cooperation across different network structures.

Li et al.'s method, while more targeted than random attacks, shows limited effectiveness, with slight increases in $\rho$ for ER and BA networks (up to around $0.15$ in ER networks) but no increase in WS networks.
This further emphasizes the superior performance of our proposed adversarial attacks in promoting cooperation across diverse network topologies.
Note that in Li et al.'s method, link weights are incrementally increased or decreased, but it has been confirmed that the change in link weights reaches $\epsilon$ in early time steps and that the original conclusions remain valid when using the average perturbation magnitude as the basis for comparison.

The trend of $\rho$ increase due to adversarial attacks varies among network structures, with BA networks showing an almost linear increase, while ER and WS networks exhibit a delayed increase, particularly pronounced in WS networks.
This is likely due to the cooperation-suppressing effect of WS networks compared to other networks, as well as the $\epsilon$ being too small.
Indeed, in previous studies, relatively large $\epsilon$ values were set to sufficiently promote cooperation.

These results suggest that adversarial attacks can promote cooperation more efficiently.
However, note that the trend of $\rho$ increase due to adversarial attacks varies somewhat depending on the network structure.
For example, in BA networks, the increase in $\rho$ is almost linear with respect to the increase in $\epsilon$, whereas in ER and WS networks, there is a delay in this increase.
This delay trend is particularly pronounced in WS networks.
These differences are likely attributable to the inherent characteristics of each network model (as mentioned earlier, BA networks have an intrinsic effect of promoting cooperation, while WS networks have an effect of suppressing cooperation).

\begin{figure}[htbp]
\begin{center}
\includegraphics[width=77mm]{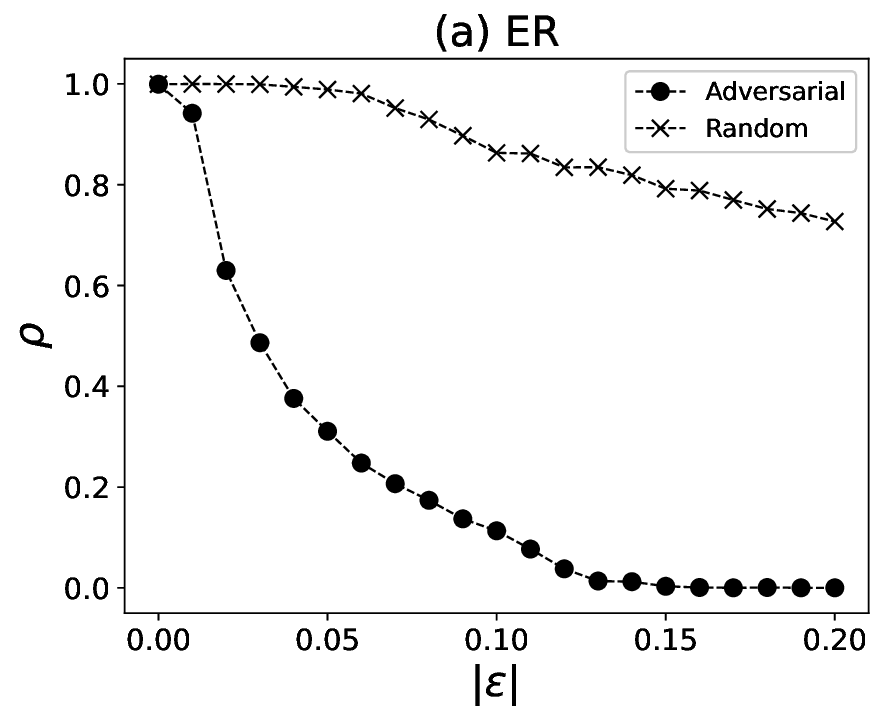}
\includegraphics[width=77mm]{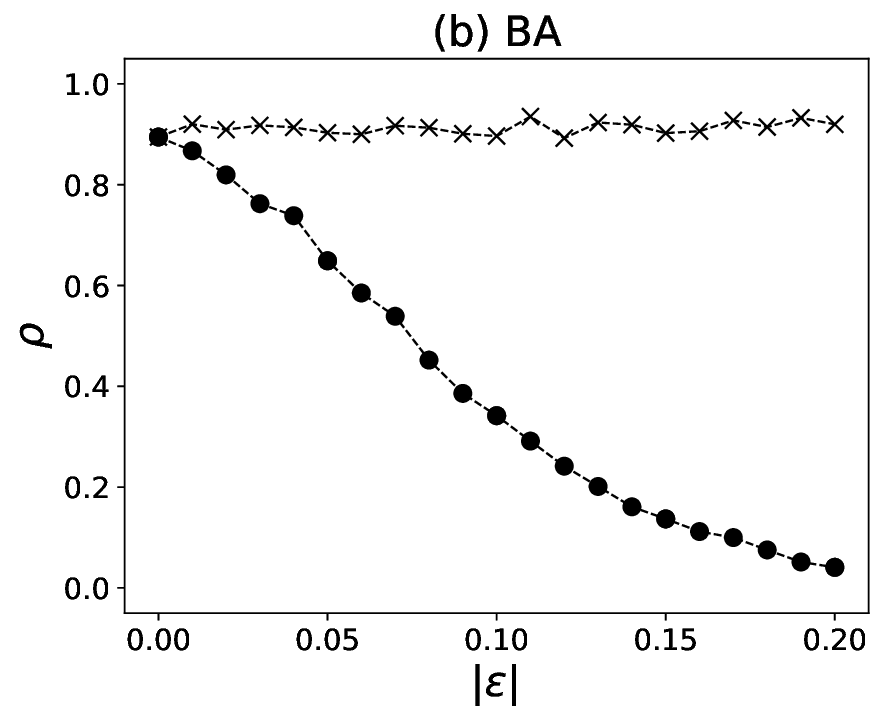} \\ 
\includegraphics[width=77mm]{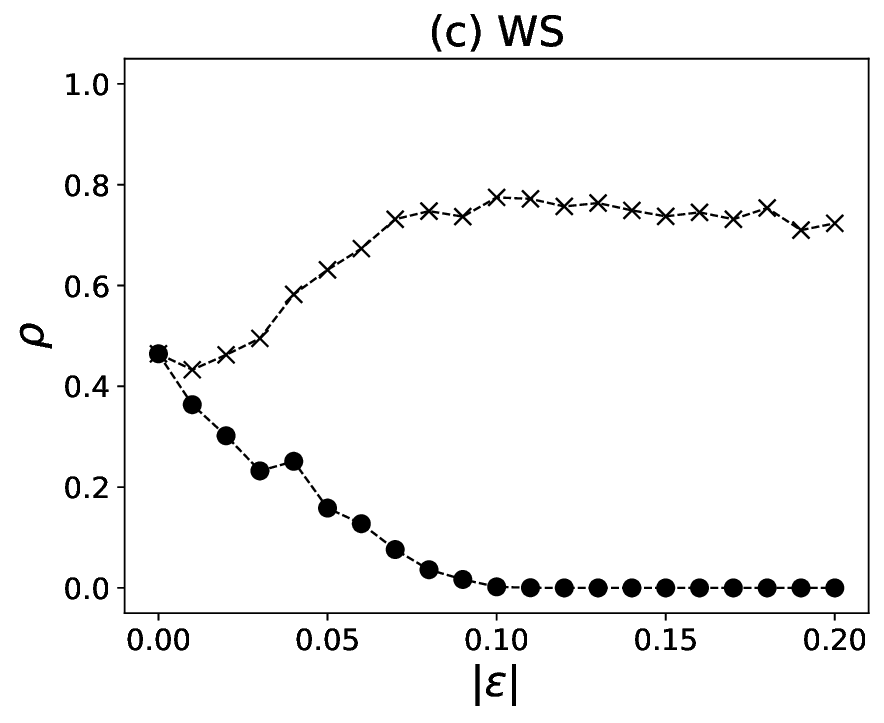}
\end{center}
\caption{\label{fig:rho_vs_eps_defect_model} Line plots of the proportion of cooperators $\rho$ versus perturbation strength $|\epsilon|$ for adversarial attacks inhibiting cooperation ($\epsilon < 0$) in (a) Erd\H{o}s--R\'enyi (ER) random networks, (b) Barab\'asi--Albert (BA) scale-free networks, and (c) Watts--Strogatz (WS) small-world networks, showing that adversarial attacks inhibit cooperation substantially more effectively than random attacks and Li et al.'s method. The advantage of defectors over cooperators $b$ is 1.1, 1.5, and 1.05 for (a), (b), and (c), respectively. These $b$ values are chosen as the maximum values for which $\rho \geq 0.9$ when $\epsilon = 0$ in Fig. \ref{fig:rho_vs_b_wrt_eps_model}, or set to 1.05 if no such value exists within $1 < b < 2$ (as in Fig. \ref{fig:rho_vs_b_wrt_eps_model}c).}
\end{figure}

For inhibiting cooperation ($\epsilon < 0$) (Figure~\ref{fig:rho_vs_eps_defect_model}), adversarial attacks consistently and effectively suppress cooperation across all model networks.
Note that Li et al.'s method is not included in this comparison because their method was specifically designed to promote cooperation.
Therefore, their method cannot be directly applied to scenarios where cooperation inhibition is desired. 
In contrast, random attacks show limited or even counterproductive effects, particularly in WS networks where they tend to increase $\rho$, likely due to the weakening of cooperation-inhibiting shortcuts.
The inhibitory effect of adversarial attacks, while consistent, is slightly influenced by network structure.
ER and WS networks show rapid decreases in $\rho$ with increasing $|\epsilon|$, whereas the decrease is more gradual in BA networks, again reflecting their intrinsic cooperation-promoting nature.

It is worth noting that promoting defection is generally easier than promoting cooperation in the prisoner's dilemma game.
This asymmetry is fundamentally rooted in the game's structure, where defection is each player's rational choice, as mentioned in the introduction.
Our results confirm this theoretical expectation, as seen in the more rapid decrease of $\rho$ when $\epsilon < 0$ (Figure \ref{fig:rho_vs_eps_defect_model}) compared to its increase when $\epsilon > 0$ (Figure \ref{fig:rho_vs_eps_coop_model})

\begin{figure}[htbp]
\begin{center}
\includegraphics[width=78mm]{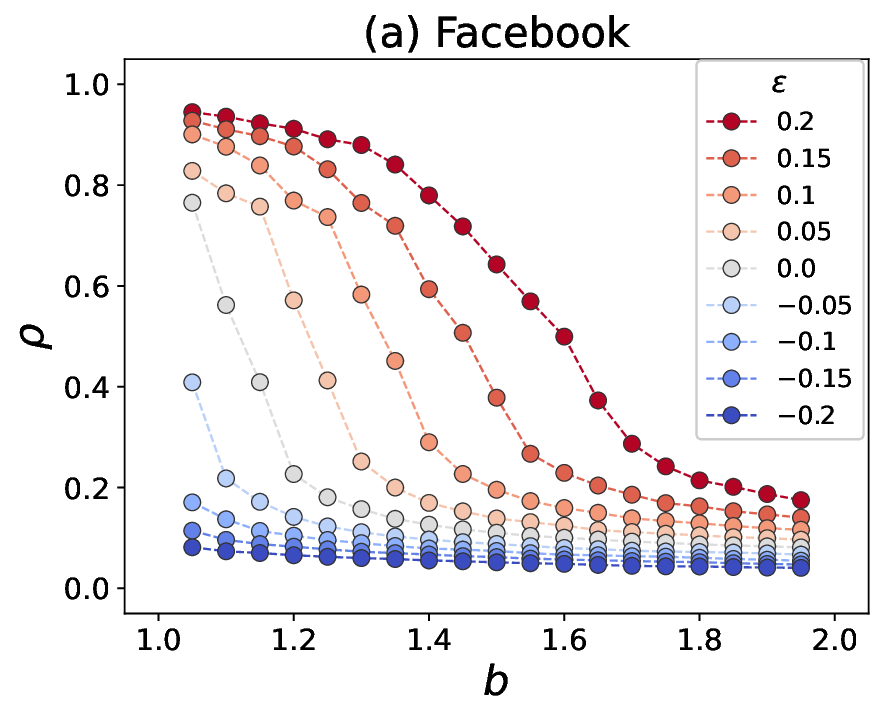}
\includegraphics[width=78mm]{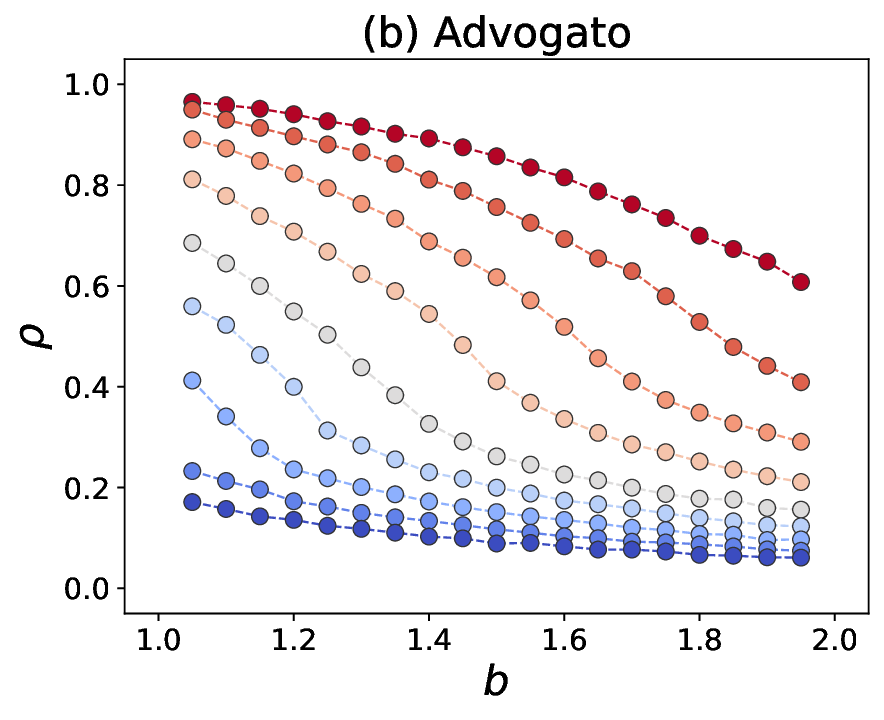} \\
\includegraphics[width=78mm]{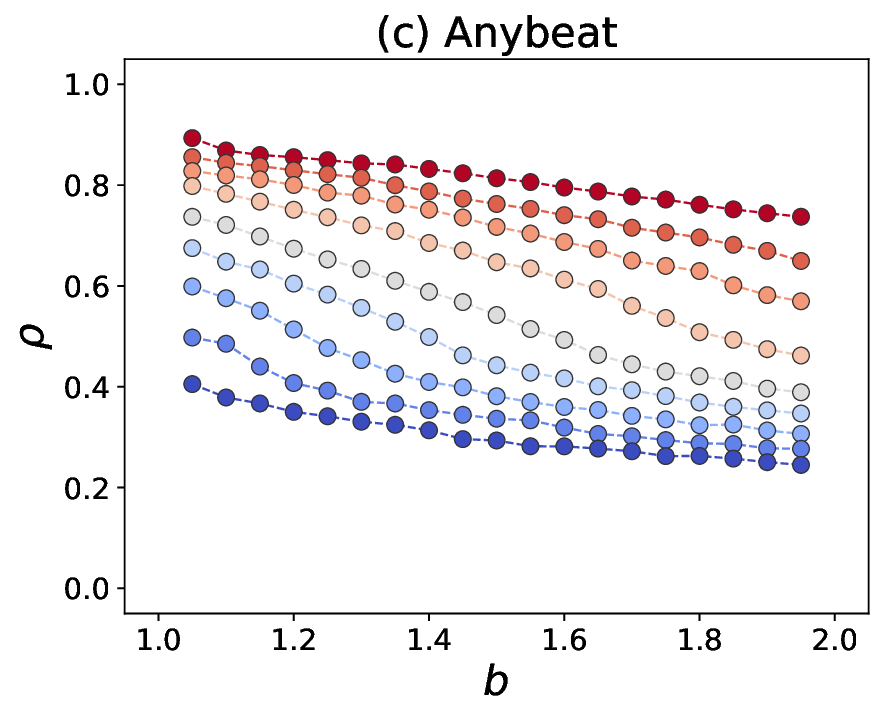}
\includegraphics[width=78mm]{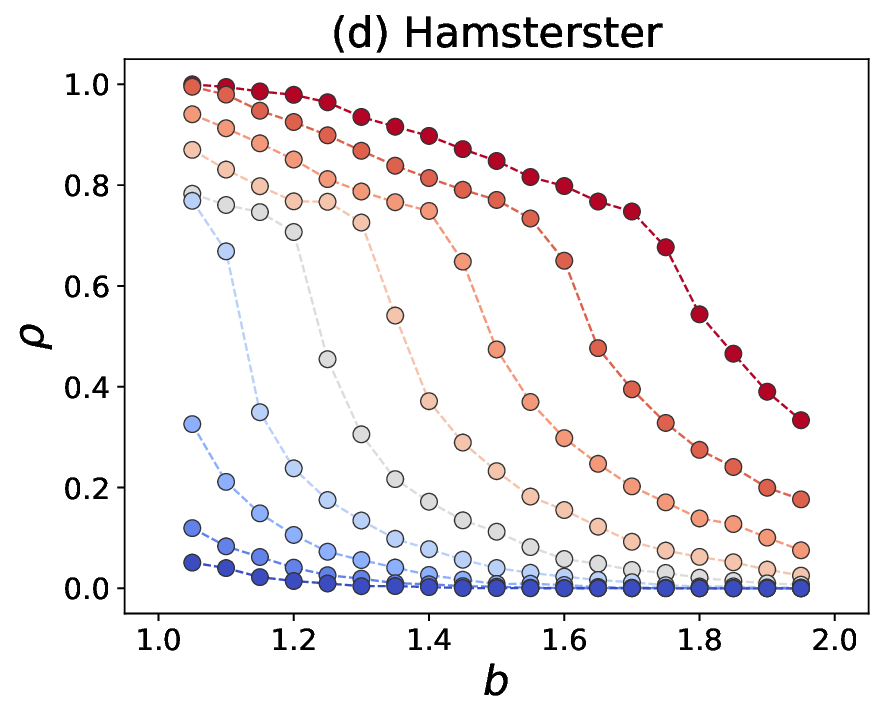}
\end{center}
\caption{\label{fig:rho_vs_b_real} Line plots of the proportion of cooperators $\rho$ versus the advantage of defectors over cooperators $b$ for different $\epsilon$ in (a) Facebook ($N=4039$ and $\langle k\rangle = 43.7$), (b) Advogato ($N=5054$ and $\langle k\rangle = 16.6$), (c) AnyBeat ($N=12645$ and $\langle k\rangle = 7.8$), and (d) HAMSTERster networks ($N=2000$ and $\langle k\rangle = 16.1$), showing that adversarial attacks ($\epsilon$) can both promote ($\epsilon > 0$) and inhibit ($\epsilon < 0$) cooperation.}
\end{figure}

The effect of adversarial attacks is further evaluated using real-world social networks (Figure \ref{fig:rho_vs_b_real}): Facebook \cite{leskovec2012learning}, Advogato \cite{rossi2015network,massa2009bowling}, AnyBeat \cite{rossi2015network,fire2013link}, and HAMSTERster networks \cite{rossi2015network}.
These networks reflect actual human relationships obtained from online social media platforms and web services, and possess diverse structural characteristics (e.g., high degree heterogeneity and high clustering coefficients; see Table 1 in \cite{chiyomaru2023mitigation} for details).
Note that these networks are undirected.
For simplicity, simulations used the largest connected component in each network, with all link weights set to 1.
The effect of adversarial attacks on real-world networks was found to be qualitatively consistent with the results from the aforementioned model networks.
The results are qualitatively consistent with those from model networks, demonstrating both promotion ($\epsilon > 0$) and inhibition ($\epsilon < 0$) of cooperation. 
However, the AnyBeat network shows comparatively smaller effects due to its sparse nature, as the method's effectiveness is limited by its constraint of modifying only existing link weights.

\section{Discussion}
This study investigated the impact of strategic manipulation of link weights, applying the concept of adversarial attacks, on the evolution of cooperation in the prisoner's dilemma game in complex networks. 
It revealed that even slight perturbations can effectively promote or inhibit cooperation (Figures \ref{fig:rho_vs_b_wrt_eps_model}--\ref{fig:rho_vs_b_real}).
This effect was confirmed not only in model networks (Figure \ref{fig:rho_vs_b_wrt_eps_model}) but also in real-world social networks (Figure \ref{fig:rho_vs_b_real}), demonstrating the versatility and robustness of adversarial attacks.
Conventional approaches, especially link weight adjustment methods represented by Li et al.'s technique, required relatively large changes in link weights to promote cooperation.
In contrast, the approach utilizing the concept of adversarial attacks proposed in this study enables effective evolution of cooperation with minimal perturbations (Figure \ref{fig:rho_vs_eps_coop_model}). 
This point is particularly crucial because, in real-world social networks, the structure and link weights are determined by individuals' intrinsic properties and relationships, making large changes unrealistic and only small modifications permissible.

These findings suggest that large heterogeneity in link weights \cite{du2008evolutionary,meng2016interdependency,iwata2016heterogeneity,hong2010evolutionary,buesser2012supercooperation,li2017co} is not necessary to promote cooperation.
The proposed adversarial attack method demonstrates that cooperation can be effectively promoted with minimal weight modifications.
This provides a new perspective on link weights' role in cooperation evolution, suggesting that the ability to fine-tune network interactions may be more crucial than static weight heterogeneity.

Link weights correspond to factors such as frequency of contact or level of trust between individuals \cite{iwata2016heterogeneity,cao2011evolution,du2008evolutionary,buesser2012supercooperation}.
In practice, implementing adversarial attacks on these weights in social networking services (SNS) can be achieved without requiring users to consciously adjust their relationships.
SNS operators with access to network data and user behavior patterns can influence social behavior by manipulating the digital environment.
This can be done through various mechanisms within the platform's architecture. 
For instance, operators can adjust content visibility in user timelines \cite{kramer2014experimental}, which alters perceived interaction frequencies.
They can also modify recommendation algorithms to promote specific types of interactions \cite{santos2021link}, implement internal trust scores that influence user perceptions and interactions, and strategically place interaction prompts or nudges to reinforce particular connections.
These modifications, based on individual behavior patterns and payoff information (Equation (\ref{eq:gradient})), can effectively alter link weights and strengthen targeted links within the network.
The minimal perturbations required, as demonstrated in our results (Figure \ref{fig:rho_vs_eps_coop_model}), make this method particularly suitable for real-world applications where large-scale changes to network structures are often unfeasible or undesirable. 

However, caution is necessary regarding the effectiveness of this adversarial attack in inhibiting cooperation (promoting defection; Figure \ref{fig:rho_vs_eps_defect_model}).
The observed asymmetry between promoting cooperation and defection aligns with the fundamental nature of the prisoner's dilemma.
Defection is the dominant strategy in this game, making it easier to push the system towards a defection-dominated state.
This explains why our adversarial attacks appear more effective in inhibiting cooperation (promoting defection) than in promoting cooperation.
However, the fact that our method can still effectively promote cooperation, despite defection being the dominant strategy, underscores its potential as a powerful tool for influencing collective behavior in complex systems.
Operators or malicious third parties with access to social network data and user behavior patterns could potentially suppress the evolution of cooperation and cause social disruption without users' knowledge. 
This point suggests a new vulnerability in digital society.
Furthermore, the application of this method extends beyond social media to various social interaction settings, such as organizational communication tools and online learning platforms. This presents both opportunities for promoting cooperative behavior and ethical challenges concerning individual autonomy. These findings have important implications for the design and operation of social systems, highlighting the need for careful ethical consideration and appropriate regulation in the use of these technologies.

The adversarial attack concept is expected to continue playing a crucial role in developing approaches to promote cooperation.
Previous methods for promoting the evolution of cooperation have taken various approaches: some utilize payoff-based mechanisms including reputation systems \cite{li2019reputation, liu2020link, cao2011evolution} while others focus on individual strategies (cooperation or defection) \cite{li2017co, fu2009partner,lee2018evolutionary,santos2006cooperation}.
In particular, Li et al.'s method \cite{li2019reputation} implements a reputation-based approach where link weights are adjusted based on accumulated payoffs, reflecting players' historical performance.
By introducing the adversarial attack concept, this study reframes these different approaches as components of an optimization problem, enabling a more systematic approach derivation.
Indeed, the method demonstrated in this study achieves more efficient promotion of cooperation by comprehensively considering both payoffs and individual strategies through the gradient calculation.
This suggests that the framework of adversarial attacks possesses the flexibility to unify and extend various existing approaches to cooperation promotion.

Further development of this approach may lead to more efficient and sophisticated methods for promoting cooperation.
For instance, the current study has constraints such as approximating the prisoner's dilemma game with mean-field time evolution to estimate gradients, and only being able to estimate the sign of the gradient.
Due to these constraints, while the proposed method was effective for representative model networks and several real-world networks, its effect may be limited for specific network structures or under dynamic conditions.
Future research directions include developing more accurate gradient estimation methods.
The exploration of more precise gradient estimation approaches that go beyond mean-field approximation is anticipated.
Moreover, developing sparse attack strategies that work effectively with manipulation of fewer links is another important challenge.
This could lead to reduced computational costs and easier implementation.

Furthermore, the adversarial attacks examined in this study are limited to complex networks where relationships between individuals are bidirectional and all link weights are equal.
Therefore, investigating adversarial attacks in complex networks where interpersonal relationships are asymmetric \cite{kim2002dynamic,gao2022asymmetric} and link weights vary significantly \cite{du2008evolutionary,meng2016interdependency,iwata2016heterogeneity,hong2010evolutionary,buesser2012supercooperation} is of great interest from the perspective of application to more realistic social network models.
In this regard, devising adaptive attack methods that can respond to temporal networks \cite{fu2009partner,li2020evolution}, multiplex networks \cite{gomez2012evolution,takesue2021symmetry,wang2015evolutionary}, and higher-order networks \cite{civilini2024explosive,ma2024social} is also an important research challenge, as real-world social networks are not static but constantly changing.

Moreover, while it has been confirmed that the conclusions remain valid when using alternative discrete update rules such as proportional imitation instead of the Fermi function, further investigation of strategy update mechanisms \cite{SZABO200797,PERC20171} could provide additional insights.

Applying these methods to different game-theoretic models \cite{SZABO200797,PERC20171} should be a key priority for future research.
While our study focused on the weak prisoner's dilemma (boundary game) for its analytical tractability, extending the analysis to the donor-recipient game, which is considered the standard template in theoretical biology \cite{wang2015universal}, would be particularly valuable.
This more general form of the prisoner's dilemma could provide additional insights into the universality of our findings.
Beyond the prisoner's dilemma variants, investigation should extend to other fundamental social dilemmas such as public goods games \cite{santos2008social,gomez2011disentangling}, coordination games \cite{broere2017network,raducha2023evolutionary,kobayashi2023dynamics}, or situations where multiple games coexist \cite{szolnoki2014coevolutionary,venkateswaran2019evolutionary}. This could deepen our understanding of cooperation promotion and inhibition mechanisms in various social dilemma situations.

Additionally, it is important to verify these methods in situations with players having diverse strategies \cite{santos2008social,nowak1993strategy}, such as the Rock-Paper-Scissors game \cite{roy2024eco} where cyclical dominance between strategies creates complex dynamics.
Beyond these basic cooperative scenarios, recent research has demonstrated the importance of considering more complex moral behaviors and reciprocity mechanisms \cite{capraro2021mathematical,xia2023reputation}.
In real societies, individuals make decisions based on moral preferences and reputation concerns. Examining the effects of adversarial attacks under such complex conditions could provide insights into how network-based interventions influence broader patterns of social behavior and moral decision-making.

Moreover, verification in social networks composed of actual humans is essential \cite{traulsen2023future,gracia2012heterogeneous,rand2011dynamic}.
By examining the extent to which findings from simulations and theoretical models are applicable to real human behavior, the practicality and social impact of this research can be evaluated.

Advancing research on adversarial attacks in evolutionary games has the potential to enhance our understanding of cooperation and competition dynamics in complex social systems, contributing to more effective social system design.

\section{Conclusions}
This study introduces a novel approach to modulating cooperation in complex networks through adversarial attacks on the prisoner's dilemma game. The research demonstrates that minimal, strategic perturbations can significantly influence cooperation evolution across diverse network structures, challenging the necessity of large link weight heterogeneity. The method's efficacy in both theoretical and real-world networks highlights its potential for practical application. By framing cooperation promotion as an optimization problem, the study offers a more systematic approach with implications for social system design (in particular, online). Future research should explore efficient strategies for varied network topologies, extend to other evolutionary games, and validate results empirically. This work advances our understanding of cooperation dynamics and provides a tool for shaping collective behavior, while emphasizing the importance of ethical considerations in digital social systems.

\section*{Data availability statement}
The data and code that support the results presented in this study are openly available in the author's GitHub repository (github.com/kztakemoto/AdvGame).

\ack
This study was supported by JSPS KAKENHI (Grant Number 21H03545).

\section*{References}
\bibliography{iopart-num}

\end{document}